\documentstyle[aps,twocolumn]{revtex}
\begin{document}

\title{Log-normal Distribution of Level
Curvatures in the Localized Regime: analytical verification}

\bigskip
\author{Michail Titov$^{1}$,
Daniel Braun$^{2}$ and Yan V.~Fyodorov$^{1,2}$}

\smallskip

\address{$^1$ Petersburg Nuclear Physics Institute,
Gatchina 188350, Russia}

\address{$^2$ Fachbereich Physik, Universit\"at-GH Essen,
D-45117 Essen, Germany}

\maketitle
\bigskip

\begin{abstract}
 We study numerically and analytically the moments of the dimensionless
level curvature for one-dimensional disordered rings of the
circumference $L$ pierced by a magnetic flux $\phi$. The negative
moments of the curvature distribution can be evaluated analytically in
the extreme localization limit.  The ensuing small curvature
asymptotics of the corresponding distribution has a "log-normal"
behavior.  Numerically studied positive moments show differences from
other log-normally distributed quantities.
\end{abstract}
\vfill

Recently there was a considerable interest in various
statistical
characteristics of spectra
of disordered and chaotic
quantum systems, see e.g.\cite{AlSi}.
One particularly
interesting issue is the so-called "level response
statistics" characterizing sensitivity of
individual energy levels
$E_n(\lambda)$
with respect to perturbation in an external parameter
$\lambda$.
The role of such a parameter can be played by e.g. an
external
electric or magnetic field, the strength and form of the
potential,
or any other appropriate parameter
 of different nature
on which the system Hamiltonian is dependent.
 A convenient
quantitative measure of
level sensitivity is provided by a set of
first and second derivatives
$\partial E_n(\lambda)/\partial\lambda$
and
$\partial^2 E_n(\lambda)/\partial\lambda^2$ , known as
"level velocities" and "level curvatures", respectively.

The
statistics of these quantities are mostly studied for the
systems
with completely "ergodic" eigenfunctions covering randomly,
but
uniformly all the available phase space and showing no
specific
internal structure. For systems of this type
most of their
statistical characteristics are known to be universal,
i.e.
independent on particular microscopic details, and
adequately
described
by ensembles of large random matrices of
particular
 global symmetry\cite{AlSi,Muz}. The same universality
class
comprises also weakly disordered (metallic) systems as long
as
effects of Anderson localization are
negligible\cite{AlSi,Muz,Efetov}.

The form of the level curvature distribution typical for
random matrices of various symmetry classes was
guessed by Zakrzewski and
 Delande \cite{ZD}
on the basis of the numerical data.
The analytical derivation of the Zakrzewski-Delande
distribution is due to von
Oppen\cite{Oppen}
and Fyodorov and Sommers\cite{FS}.

The effects
of eigenfunction localization are expected to modify
the level curvature
statistics drastically.
In contrast to the
"random matrix" regime it
is
obvious that the level response should be less universal
 depending on the nature of perturbation. Such a non-universality
manifests
itself already in the form of perturbative
corrections to the curvature distribution due to localization
effects \cite{Yurk}.

Actually, the sensitivity of energy
spectrum
to a change in boundary conditions was suggested long
ago
by Thouless as a measure of system
conductance\cite{Thou}.
To quantify this statement let us
consider a one-dimensional sample closed to form a
ring of circumference
$2L$ encircling
the Aharonov-Bohm flux $\phi$
(measured in units of the
flux quanta $\phi_0=c\hbar/e$). The
wavefunction in such geometry
acquires the phase when going around
the flux:
\begin{equation}\label{per}
\Psi(x+L)=e^{i\phi}\Psi(x-L).
\end{equation}
 The
curvature is given by
the second-order perturbation
theory:
\begin{equation}\label{2}
K_n=\left. \frac{\partial^2
E_n(\phi)}{\partial\phi^2} \right| _{\phi=0}
\propto\sum_{m\ne n}\frac{\mid\langle m\mid
\hat{P}_x\mid
n\rangle\mid^2}
{E_m(0)-E_n(0)},
\end{equation}
up to a constant shift.
Here
$\mid m\rangle$ and $E_m(0)$ are eigenvectors and eigenvalues of
the Hamiltonian at zero flux and $\hat{P}_x$ is the momentum
operator.
Using a similarity of this expression to the conductance
given by
the Kubo formula,
Thouless argued that the "typical"
dimensionless level
curvature $K/\Delta$
(with $K$ measured, e.g.
by the widths of the curvature distribution)
is proportional to the
dimensioneless conductance of the sample.

Thouless original
qualitative arguments seemed to be
controversial and also did not
take into account strong correlations
between energy levels of a
disordered system. The
curvature-conductance relation gave rise
to
a lot of discussion and even claimed to be
incorrect\cite{AL}.
The problem was reconsidered in much details
recently
\cite{Akk,Mont,Sano}, the results favouring validity of the
Thouless
idea in the metallic regime. Actually,
 for a disordered
system in good metallic regime
 the Thouless relation, including
the universal proportionality coefficient, was derived
explicitly \cite{FSPRE} in the course of analytical
verification of the Zakrzewski-Delande distribution.
However, the perturbative localization corrections to both quantities
are different\cite{Yurk}.

As to the issue of the level
curvatures in
systems with strongly localized eigenstates, our present
understanding is
based mainly on the results of
numerical
simulations\cite{Mont,Karol,Krav,Braun}.
It turns out that the
curvature
distribution is close to log-normal one for strong
enough
localization and has quite a non-trivial form in the vicinity
of
the Anderson transition\cite{Krav}. Qualitative origin
of the
log-normal distribution can be inferred
from
Eq.(\ref{2})
 as suggested in \cite{Krav}. Indeed, it is
natural to assume that the
absolute value of the curvature $|K|$ is proportional to
the product of amplitudes of a typical wavefunction at
opposite edges $\pm L$ of the sample: $|K|\propto |\Psi(-L)\Psi(L)|\sim
\exp{-\left(\mbox{const} L/\xi\right)}$. The log-normal distribution
follows if one assumes that the inverse localization
length $\xi^{-1}$ shows Gaussian eigenstate-to-eigenstate
fluctuations.

Whatever attractive and transparent is this simple argumentation,
a close inspection shows that it should be taken with caution.
To this end let us recall that
the distribution of the quantity $v=|\Psi(-L)\Psi(L)|$
for (quasi) one dimensional disordered samples can be calculated
 analytically \cite{FMJETP}. It indeed has a log-normal form
for $L\gg \xi$, but all the positive moments $\langle v^n\rangle$
are dominated by rare events in such a way that $\langle v^n\rangle
\propto \exp{-c L/\xi} $, where the constant $c$ is {\it independent}
of the index $n$.

To check if this property is shared by the level curvatures
 we performed numerical
simulations of the tight--binding
Anderson model in one dimension. The
curvatures were calculated exactly from both eigenvalues and eigenvectors
(see \protect\cite{braun}), between 100 and 500 disorder realisations
were used. We plot the numerical results for the logarithm of
first, second and third moments of the curvature distribution
versus $L$, see fig.1.
The moments indeed decay roughly exponentially with system size, but
the typical decay length decreases for higher moments in contrast to
the behaviour typical for the quantity $|\Psi(-L)\Psi(L)|$.

The second fact to be mentioned is that
 earlier numerical investigations discovered
a peculiar feature of the curvature distribution: the log-normal
law is quite a good fit separately for the domain of curvatures
larger and smaller than the most probable value $\langle
\ln{|K|}\rangle \sim
L/\xi$, but  the parameters of the fit are slightly different for
the two domains. At the same time, the statistics of the
correlation function
$v=|\Psi(-L)\Psi(L)|$ is truly log-normal everywhere \cite{FMJETP}.

 These observations provide us with a motivation to
consider the problem of the level curvature distribution
in strongly localized regime on a more
sound basis without invoking any additional assumptions.

In the present paper we treat
analytically the problem of the
distribution of absolute value of
the level curvatures:
\begin{equation}\label{3}
{\cal
P}(K)=\Delta\left
\langle\sum_{n}\delta(E-E_n(0))\delta(K-\mid
K_n\mid)\right\rangle
\end{equation}
 for a one-channel ring
characterized by the Schr\"{o}dinger
equation:
\begin{equation}\label{4}
\left(-\frac{d^2}
{dx^2}+U(x)-E\right)\Psi(x)=0
\end{equation}
with
the boundary condition Eq.(\ref{per}).
Here $U(x)$ is a white
noise
potential $\langle U(x)U(x')\rangle=D\delta(x-x')$
which is
considered to be weak: $l=4k^2/D\gg k^{-1}$,
with $k$ being the
Fermi momentum related to the energy $E$ as $k^2 = E$ and $l$
standing for the mean free
path. For the
one-channel ring the mean free path is of the same order
as the
localization length.  As we shall see,
 the negative moments of the curvature
distribution Eq.(\ref{3}) can be
found explicitly for rings with
$L\gg l$, i.e. in strong localization
regime. This allows us
to verify analytically the log-normal
 nature of the ensuing
distribution at small curvatures.

To address the problem
of level curvatures in the most direct way
we follow the method
by Dorokhov\cite{Dor} and Kolokolov\cite{Kol} who calculated the
absolute value of the persistent current $j_n=\mid
\frac{\partial
E_n}{\partial \phi}\mid$. To this end, let us
 associate with any point $x;\quad -L\leq x \leq L$ within the
sample a vector ${\bf V}(x)=\left(\begin{array}{c}v_{+}(x)\\
v_{-}(x)\end{array}\right)$ with components:
$v_{\pm}(x)=\pm(d\Psi/dx\pm ik\Psi)\exp{(\mp ikx)}$ and
consider a $2\times 2$
transfer matrix ${\cal T}$ relating the
value ${\bf V}(x)$ to "initial" value ${\bf V}(-L)$
in the following way: ${\bf V}(x)=\hat{\cal
T}(x,-L){\bf V}(-L)$.
Due to the current conservation
 the transfer
matrix can be parametrized as
\begin{equation}\label{6}
\hat{\cal
T}(x,-L)=\left(\begin{array}{cc}\cosh{\Gamma}e^{i\alpha}&
\sinh{\Gamma}e^{i\beta}\\ \sinh{\Gamma}e^{-i\beta}&
\cosh{\Gamma}e^{-i\alpha}.
\end{array}\right)
\end{equation}
Here $\alpha(x,E),\beta(x,E)$
and $\Gamma(x,E)$ are real functions to be determined.
The
periodic boundary condition Eq.(\ref{per}) can be written
in terms of the transfer matrix
$\hat{\cal T}=\hat{\cal T}(L,-L)$
as $\mbox{det}\left(\hat{\cal
T} e^{2(ikL\hat{\sigma}_z)}-e^{i\phi}\right)=0$, where
$\hat{\sigma}_z$ stands
for the Pauli matrix. It is convenient
to rewrite this condition in terms of the
$\hat{\cal T}$- matrix elements as:
\begin{equation}\label{7}
f(E_m,L)=
\cos{\phi};\quad f(E,x)=\cosh{\Gamma}
\cos\left(\alpha+2kx\right)
\end{equation}

This
equation determines the set of energy levels $E_m(\phi)$.
The following identities can be immediately inferred from this
fact:
\begin{equation}
\label{8}
\sum_{n}\delta(E-E_n(\phi))=
\delta\left( f(E,L)-\cos{\phi}\right) \left|df/dE\right|,
\end{equation}
\begin{equation}\label{8a}
\left| \frac{\partial^2 E_n}{\partial
\phi^2}(\phi=0)\right|
=\left|\frac{\partial f}{\partial E}\left(E=E_n\right)\right|^{-1}.
\end{equation}

As a result one can write down the moments of
the distribution
Eq.(\ref{3}) in the following
form:
\begin{equation}\label{9}
M_n=\int_{0}^{\infty}K^n{\cal
P}(K)=\frac{1}{\Delta^{n-1}}
\left\langle
\delta(f-1)\left|\frac{\partial f}{\partial
E}
\right|^{-n+1}\right\rangle.
\end{equation}

When performing the
disorder averaging it is convenient to get rid of
the
$\delta-$functions in Eq.(\ref{9}) by averaging the
corresponding
expression over the ensemble of samples with slightly
fluctuating
sample lengths $L\pm \delta L;\quad k^{-1}\delta L\ll
\min(l,L)$
\cite{Kol}. When doing this we take into account
 that the functions
$\Gamma(x,E),
\alpha(x,E)$ change very slowly on the scale $\delta x
\sim
k^{-1}$, i.e.
$(d\Gamma/dx,d\alpha/dx)\ll k$.
As a consequence
 there is
typically only one solution of the equation Eq.(\ref{7}) in the
interval $\delta x=\pi/k$.
We also found that the quantities $\frac{d\alpha}{dE}, Lk^{-1}$
are  of the lower order $O(1/(Lk))$ when compared with
$\frac{d\Gamma}{dE}$ and can be safely neglected.
 This gives the possibility to rewrite
the expression Eq.(\ref{9}) in the form:
\begin{equation}\label{10}
M_n\approx\frac{1}{2\pi\Delta^{n-1}}
\left\langle\frac{1}{\mid\sinh{\Gamma}\mid}
\left|\frac{d\Gamma}{dE}\tanh{\Gamma}
\right|^{-n+1}\right\rangle.
\end{equation}

The disorder averaging in expressions of such a type can be performed
by employing the functional integral method suggested by Kolokolov
\cite{Kol}. For accomplishing such a calculation it is
important to express the  quantity to be averaged
 in terms of elements of the matrices
${\cal T} \equiv
{\cal T}(L,-L)$ and $d{\cal T}/dE$ without involving complex
conjugation.
To find such a representation let
us consider
the auxiliary quantity:
\begin{equation}\label{11}
A=\left( 0,1 \right) \> {\cal T}s^{-}s^{+}{\cal T}
\left(
\begin{array}{c} 0 \\1\end{array}\right),
\end{equation}
where $s^{\pm}$ are familiar lowering and raising operators for the
spin $1/2$.
Taking the parametrization
Eq.(\ref{6}) into account we find that $ A= \cosh^{2}{\Gamma}
e^{-2i\alpha} $ and correspondingly
\begin{equation}\label{12}
\frac{dA/dE}{2A} =
\frac{d\Gamma}{dE} \tanh{\Gamma} \!
-i\frac{d\alpha}{dE}=
\frac{d\Gamma}{dE} \tanh{\Gamma}\left(\! 1+O \!
\left(\! \frac{1}{Lk}\! \right) \right)
\end{equation}
The right hand side of this expression is just the important
part of the combination
appearing in the expression Eq.(\ref{10}) under the sign of averaging.

 From the other hand the same
quantity $\frac{dA/dE}{2A}$
can be written in terms of the matrix $d{\cal T}/dE$ when
differentiating Eq.(\ref{11})
with respect to E.

The matrix $d{\cal T}/dE$ itself
 can be determined from
the following exact relation
between the solution $\Psi(x)$ of the
initial Schr\"{o}dinger equation
and its derivative with respect to the energy
$\Phi(x) = d\Psi(x)/dE $.
\begin{equation}\label{13}
\Phi(x) =
\Psi(x) \left[
c_{1}- \int_{-L}^{x}\frac{dy}{\Psi^{2}(y)}
\left(
c_{0}+ \int_{-L}^{y}dy_{1}
\Psi^{2}(y_{1})\right)
\right]
\end{equation}
where
$c_{0}=
\left( \Phi \> d\Psi/dx - \Psi \> d\Phi/dx
\right)|_{x=-L} $, and
$c_{1}={\Phi(-L)}/{\Psi(-L)}$.

We can write the relation:
\begin{equation}\label{14}
\frac{d{\cal T}}{dE}{\bf V}(-L)={\bf F}(L)-{\cal T}{\bf F}(-L),
\end{equation}
where ${\bf F}(x)=\frac{d{\bf V}(x)}{dE}$.
The components  $f_{\pm}$
of the vector ${\bf F}$ are represented as
integral functionals of the
fields $v_{\pm}(x)$, the latter fields having boundary values
$v_{\pm}(-L)$ at the point $x=-L$.

Combining all this facts together one can find the appropriate
representation for the quantity $\frac{dA/dE}{2A}$.
 Skipping the cumbersome intermediate steps
in favour of presenting the final expression we find finally:
\begin{equation}\label{15}
\frac{dA/dE}{2A} \simeq
\int_{-L}^{L}\frac{dy}{v_{+}(y)v_{-}(y)}
\int_{-L}^{y}v_{+}(y_{1})v_{-}(y_{1})dy_{1},
\end{equation}
where the components $v_{\pm}(x)$
of the field ${\bf V}(x)$ are given
by $
{\bf V}(x) = {\cal
T}(x,-L)\left(\begin{array}{c} 0 \\
1\end{array}\right)$.
In the expression Eq.(\ref{15}) we kept only the leading terms
with respect to $1/Lk$ and omitted also
all terms containing the fast oscillating factors $e^{\pm 2ikL}$.

The following comment is appropriate here.
Generally speaking, the
functions $v_{\pm}(x)$ are random and can take any
complex values.
However, they are not
independent from each other and can be chosen
to satisfy the constraint
$v_{+}(x)=v_{-}^{*}(x)$.
When using the Kolokolov's method of
averaging the
random fields, $v_{+}(x)$ and $v_{-}(x)$
should be analytically continued on the surface determined by this
constraint to provide the convergence of the
corresponding path integral. Apart from this fact, in our case
one can use the same
constraint to argue that $d\alpha/dE$ is small in comparison with
$\tanh{\Gamma}\> d\Gamma/dE$ by the factor $1/Lk$ which we
intensively used to simplify expressions given above, e.g.
Eq.(\ref{12}).

Eq.(\ref{10}) combined with Eqs.(\ref{12}-
\ref{15}) provides a representation which is used as a starting point for
employing the Kolokolov's approach. Indeed, we
expressed the moments $M_n$
 in terms of the elements of the ${\cal T}$-matrix without
complex
conjugation: the components of $v_{\pm}(x)$
 were related
to the ${\cal T}$-matrix above
 and $\mid \sinh{\Gamma} \mid$ can be represented in terms of the
${\cal T}$-matrix in the same way as in\cite{Kol}.

After the set of manipulations identical to those
used in \cite{Kol} we represent  the moments in the form
of the following path
integral:
\begin{equation}\label{20}
\begin{array}{l}
M_n=\frac{2}{\pi
a\Delta^{n-1}}e^{-\frac{aL}{4}}
\int\limits_{-\infty}^{\infty}d\sigma
d\sigma'

\hspace*{-0.80cm}
\int\limits_{\qquad \xi(-L)=\sigma' ;\>
\xi(L)=\sigma}
\!\!{\cal D}\xi e^{-\frac{\xi(-L)}{2}} \times
\\ \exp{\left\{
-\frac{1}{2a}\int\limits_{-L}^{L}dx \left({{\xi}}^{2} + e^{-\xi}
\right) \right\} } \>
\left[\int\limits_{-L}^{L}e^{-\xi(x)}dx
\int\limits_{-L}^{x}e^{\xi(y)}dy
\right]^{-n+1}
\end{array}
\end{equation}
where $a \equiv 2/l$.

In general, only the first moment $(n=1)$ related to
the
average persistent current  can be explicitly
evaluated for arbitrary relation between the mean free path $l$ and
the ring
circumference $2L$ \cite{Dor,Kol}.

However, in the limit  $2L \gg l$  we can
extract the
 leading contribution to the path integral Eq.(\ref{20})
for a large, but limited number of negative moments satisfying $\mid
\!n\!\mid
\> \lesssim aL/2 $.
To see this let us use the notation $m=-n$
so that the twofold integral in Eq.(\ref{20}) is
raised to the
positive power $m+1$.
It gives a possibility to treat the
path integral in
Eq.(\ref{20}) as the sum of $2(m+1)$-fold integrals
from the
corresponding matrix elements. To illustrate the structure of
the terms we write down explicitly the simplest one:
\begin{equation}
\begin{array}{l}
M_0=
\frac{2}{\pi
a\Delta^{n-1}}e^{-\frac{aL}{4}}\int_{-L}^{L}dx\int_{-L}^x dy\times\\
\left\langle 1\left|e^{-(L-x)\hat{\cal H}} t^2 e^{-(x-y)\hat{\cal H}}
\frac{1}{t^2} e^{-(y+L)\hat{\cal H}}\right|t\right\rangle\end{array}
\end{equation}
where we introduced the new variable $t=({2}/{a})
e^{-\xi/2}$ in terms of which the Hamiltonian operator
$\hat{\cal H}$  and the scalar product $\langle f|g\rangle$
read $\hat{\cal H}=\frac{2}{a}
\left(t^2\partial_t^2+t\partial_t-t^2\right)$ and
$\langle f|g\rangle=\int_0^\infty\frac{dt}{t} f g$, correspondingly.

Higher moments will contain similar integrals over
variables $-L<x_1,y_1,x_2,y_2,....x_m,y_m<L$.
Each integral of this type can be
evaluated analytically if we expand powers of $t$ in terms of
the eigenfunctions of the operator ${\cal H}$
which  are the modified
Bessel functions:${\cal
H}{ K}_{p}(t)= -{a}/{8} p^{2}{ K}_{p}(t)$.
The functions with imaginary indices $p=i\nu$ form a complete
orthogonal set suitable for expansion.
In this way we find that the leading contribution to the
path integral Eq.(\ref{20})in the limit $L\gg l$
 corresponds to the configuration:
$y_1\approx y_2\approx ...\approx y_m=-L;\quad
x_1\approx ...\approx x_m=L$ when we can effectively
write:
that
\begin{equation}\label{23}
\begin{array}{l}
\left[
\int\limits_{-L}^{L}e^{-\xi(x)}dx
\int\limits_{-L}^{x}e^{\xi(y)}dy
\right]^{m+1} \\
\approx \left[ (m+1)!
\right]^{2}
a^{-2(m+1)}e^{-(m+1)\xi(L)}e^{(m+1)\xi(-L)}
\end{array}
\end{equation}
for
$0 \leq m \ll aL/2$. For larger $m$ the contribution from omitted
terms becomes comparable with that given by Eq.(\ref{23})
 due to large combinatoric factors.

The moments $M_{-m}$ in this approximation is
equal to:
\begin{equation}\label{24}
M_{-m}=
\frac{1}{m\sqrt{\pi}}e^{-aL/4}
\left[ (m+1)! \right]^{2}{\bf
\Gamma}(2m+\frac{3}{2})
e^{aL(m+1/2)^2}.
\end{equation}
When restoring from these moments the asymptotic
behaviour of the curvature distribution
the factorial coefficients can be omitted
in view of the condition $ m\ll aL/2$. This results in the
log-normal curvature distribution:
\begin{equation}\label{25}
{\cal P}(\ln{K}) \simeq
(4\pi
aL)^{-1/2}e^{-\frac{1}{4aL}\left[\ln{K}+aL\right]^2};
\end{equation}
which is valid
in the strongly localized limit $aL \gg 1$
 inside the parametricallly
large domain of small curvatures $
-(aL)^2 \ll \ln{K} \ll -aL
$.

In conclusion, we demonstrated explicitly that the statistics
of level curvatures in $1D$ disordered systems is log-normal
inside a parametrically large domain of curvatures smaller than
the typical value.
 At the same time, interesting issues of explaining
an unexpected difference
in behaviour
of the positive moments of the curvature and that of the
eigenfunction correlator, see fig.1, as well as the
mentioned
asymmetry of the curvature distribution remain open.

The authors are grateful to H.-J.Sommers for his valuable
critical comments, to I.Kolokolov for important clarifying
communications in the process of research
 and to V.Kravtsov for useful correspondence and interest in the work.
The financial support by Russian Foundation for Basic Researches
(Grant No. 96-02-18037a) (M.T.), program "Quantum Chaos"
(grant No. INTAS-94-2058)  (M.T. and Y.V.F), and by
SFB 237 "Unordnung and Grosse Fluktuationen"
is acknowledged with thanks.


\begin{figure}
\caption{1st, 2nd and 3rd moment of the curvature
distribution as function of the system size as obtained from numerical
simulations of the tight--binding
Anderson model in one dimension. The disorder parameter $w$ (width
of the box--distribution of the
diagonal matrix elements) is $w=2$. Typical error bars
are comparable to the remaining fluctuations on top of the exponential
decay.}
\label{mom}
\end{figure}

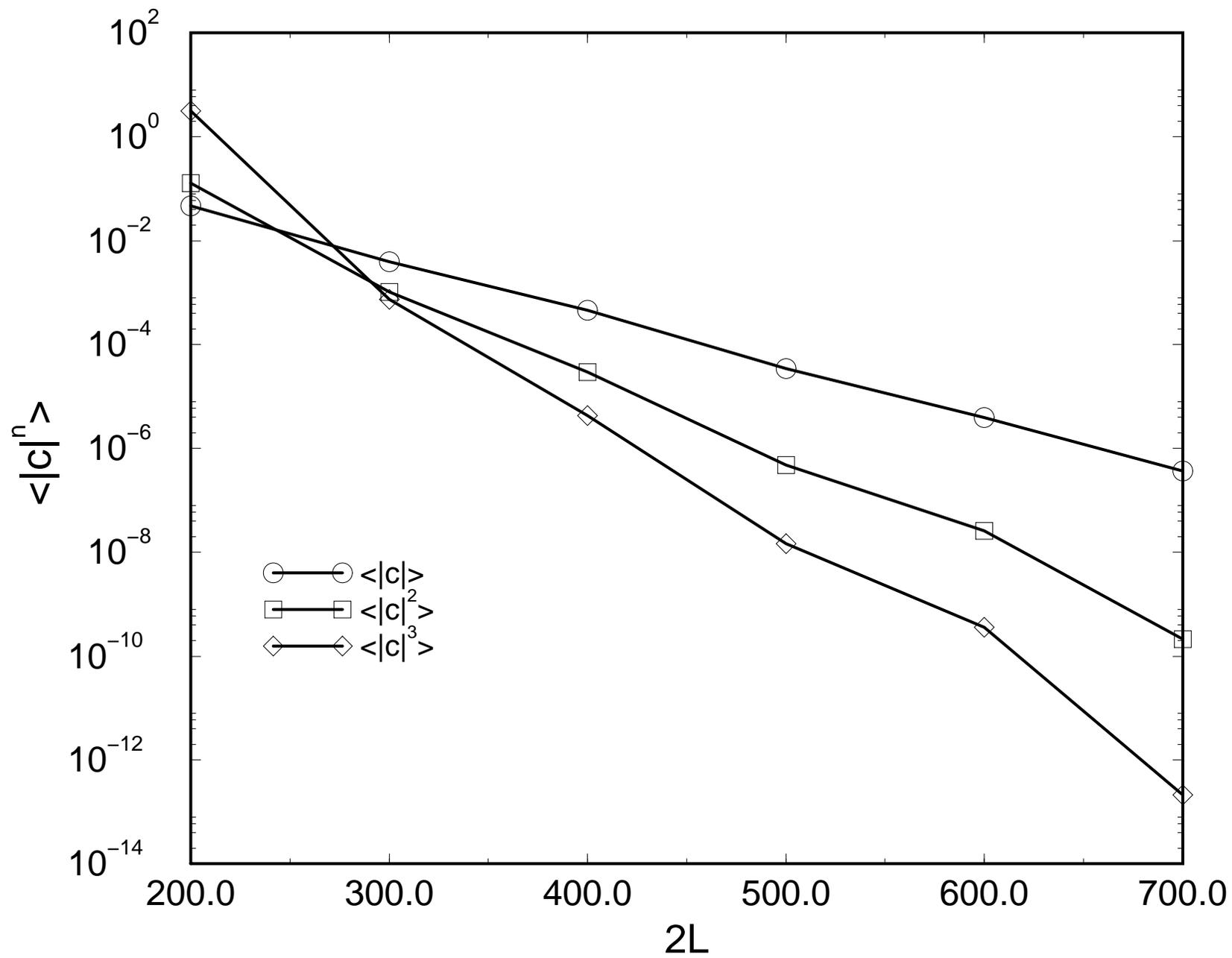
\end{document}